\documentclass[usenatbib,usegraphicx,usedcolumn]{mn2e}
\usepackage{bm}
\usepackage{amssymb,amsmath}
\usepackage{graphicx}
\usepackage{natbib}
\usepackage{program}
\usepackage{color}
\definecolor{darkgreen}{rgb}{0.15,0.5,0.15}
\definecolor{darkblue}{rgb}{0.15,0.15,0.5}

\newcommand\kpc[1]{~\rm{kpc}#1}

 \usepackage[breaklinks,colorlinks,citecolor=blue]{hyperref}
\title[Orbital pericenters and inferred DM halo structure]{Orbital pericenters and the inferred dark matter halo structure of satellite galaxies}
\author[Victor H. Robles and James S. Bullock]{Victor H. Robles\thanks{E-mail: victor.roblessanchez@yale.edu}$^{1,2}$, James S. Bullock$^{1}$ \\
$1$ Department of Physics and Astronomy, University of
       California, Irvine, 4129 Frederick Reines Hall, Irvine, CA 92697,
       USA\\
$2$ Yale Center for Astronomy and Astrophysics, New Haven, CT 06520, USA\\}

\begin{document}
%\date{MNRAS, xxx 2020}

%\pagerange{\pageref{firstpage}--\pageref{lastpage}} \pubyear{2019}

\maketitle

\label{firstpage}
%###########################################################################################################

\begin{abstract}
Using the phat-ELVIS suite of Milky Way-size halo simulations, we show that subhalo orbital pericenters, $r_{\rm peri}$, correlate with their dark matter halo structural properties.  Specifically, at fixed maximum circular velocity, $V_{\rm max}$, subhalos with smaller $r_{\rm peri}$ are more concentrated (have smaller $r_{\rm max}$ values) and have lost more mass, with larger peak circular velocities, $V_{\rm peak}$, prior to infall.  These trends provide information that can tighten constraints on the inferred $V_{\rm max}$ and $V_{\rm peak}$ values for known Milky Way satellites.  We illustrate this using  published pericenter estimates enabled by {\it  Gaia} for the nine classical Milky Way dwarf spheroidal satellites.   The two densest dSph satellites (Draco and Ursa Minor) have relatively small pericenters, and this pushes their inferred $r_{\rm max}$ and $V_{\rm max}$ values lower than they would have been without pericenter information.  For Draco, we infer $V_{\rm max} = 23.5 \, \pm 3.3$ km s$^{-1}$ (compared to $27.3 \, \pm 7.1$ km s$^{-1}$ without pericenter information). Such a shift exacerbates the traditional Too Big to Fail problem. Draco's peak circular velocity range prior to infall narrows from $V_{\rm peak} = 21 - 49$ km s$^{-1}$ without pericenter information to $V_{\rm peak} = 25-37$ km s$^{-1}$ with the constraint.  Over the full population of classical dwarf spheroidals, we find no correlation between $V_{\rm peak}$ and stellar mass today, indicative of a high level of stochasticity in galaxy formation at stellar masses below $\sim 10^7$ M$_\odot$. As proper motion measurements for dwarf satellites become more precise, they should enable useful priors on the expected structure and evolution of their host dark matter subhalos.
\end{abstract}

%\pacs{ 67.10.-j, 67.25.dk, 47.37.+q }

\section{Introduction}\label{sec:intro}

The cold dark matter (CDM) paradigm shows excellent agreement with the observed cosmological evolution \citep{springel16,vogel14a}. However, below the Local Group scale ($\leq 1 \rm Mpc$) there are potential mismatches such as the discrepancy between the number of observed and expected satellite galaxies \citep{klypin99,moore99,tollerud08} and the well-known Too-Big-too-Fail (TBTF) issue \citep{boylan11,boylan12,kimmel14a} where CDM subhalos are too dense to host the brightest Milky-Way (MW) dwarf galaxies. Current CDM simulations of MW-like galaxies ($M_{vir} \approx 10^{12} M_\odot$) suggest that tidal disruption of small-mass ($V_{\rm max} \leq 40 \rm km/s $\footnote{We define $V_{\rm max}$ as the maximum of the circular velocity $V_{\rm max}=\rm max{[GM(<r)/r]^{1/2}}$.})  halos having multiple/strong interactions with the MW disc could help to alleviate these discrepancies \citep{onghia10,zolotov12,sawala16,wetzel16,kimmel17}. Alternatively, some works have explored tidal stripping in non-standard dark matter models \citep{robles15,robles19,dooley16,dicintio17}.

An assessment of the strength of tidal disruption in real MW dwarf galaxies requires a comparison with simulations whose baryonic potential closely matches that of our MW. 
While MW-mass hydrodynamic CDM simulations are not \textit{a priori} selected to accurately fit the MW baryonic mass distribution, the approach of including time-dependent analytical potentials in CDM simulations has proven to be successful in modeling in detail the baryonic component in our galaxy at much lower computational cost allowing a rapid exploration of different MW halo masses within observational constraints \citep{kelley18,kimmel17}. 

In this Letter, we use simulations presented first in \cite{kelley18} to explore correlations between a subhalo's pericenter radius and its past mass loss and associated density structure today.  We show that smaller pericenters correlate with having higher concentrations and more mass loss and explore how this correlation can be used to provide tighter constraints on the maximum circular velocities of the nine classical MW dwarf spheroidal galaxies (dSphs).   We rely on pericenter distances from \citet{Fritz18} derived from the most recent data given by the Gaia collaboration \cite{brown18,helmi18}.

\begin{figure}\label{fig1}
\centering
\includegraphics[scale=.16,keepaspectratio=true]{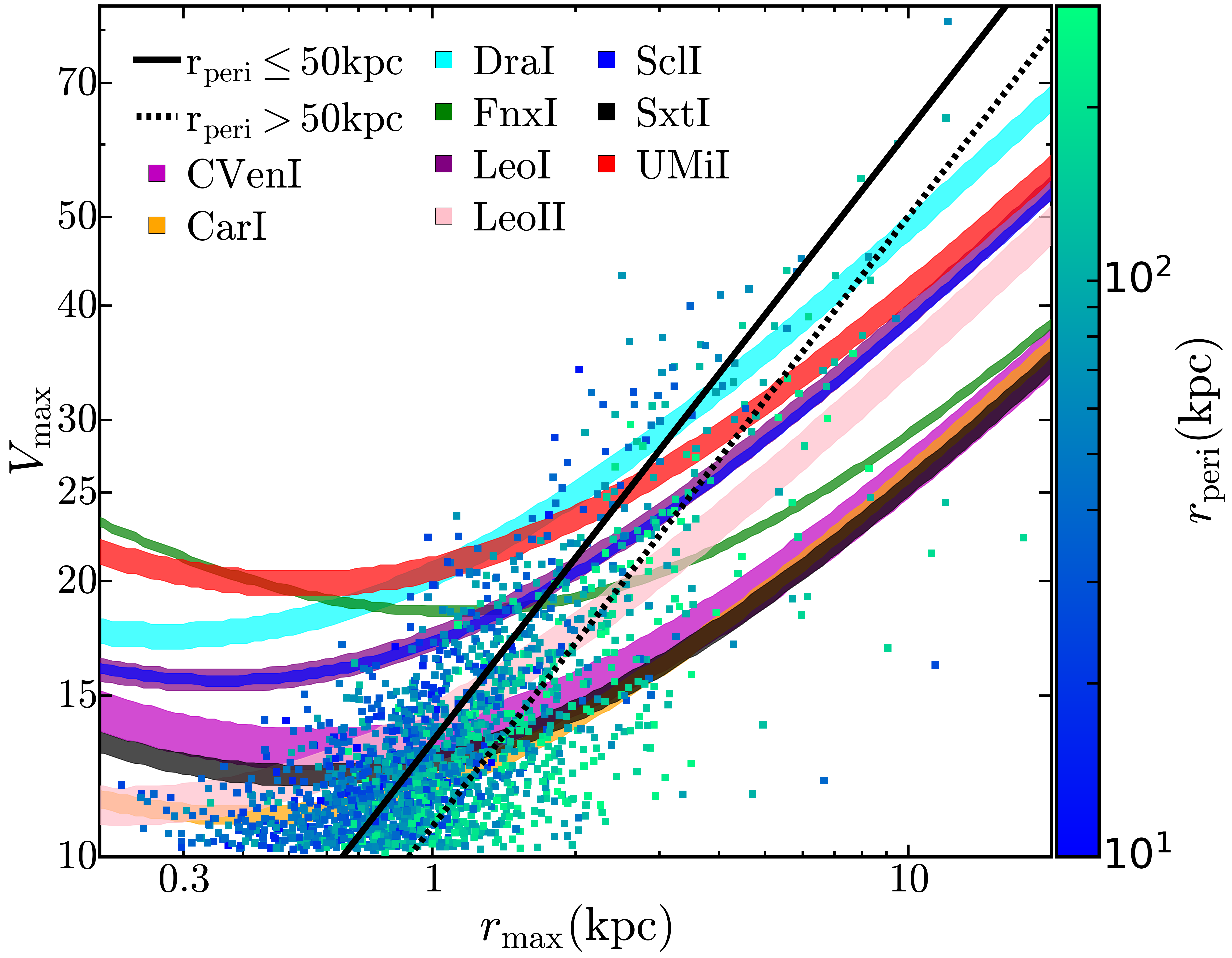} 
 \caption{
Points show $V_{\rm max}-r_{\rm max}$ values at $z=0$ for subhalos in all simulations, color coded by their pericenter radii.  Subhalos with smaller pericenters are more concentrated, with smaller $r_{\rm max}$ values at fixed $V_{\rm max}.$ The straight lines show power-law fits to the median relation for subhalos with $r_{\rm peri} \leq 50 \rm kpc$ (solid black) and $r_{\rm peri} > 50 \rm kpc$ (dashed line), respectively.  The colored bands show the 1-$\sigma$ region of parameter space allowed for the nine bright MW dwarf spheroidal galaxies ($L_{V} \geq 10^5 L_\odot$) based on their measured half-light masses assuming NFW profiles.
}
\end{figure}

\begin{figure} \label{fig2}
\centering
\includegraphics[scale=.166,keepaspectratio=true]{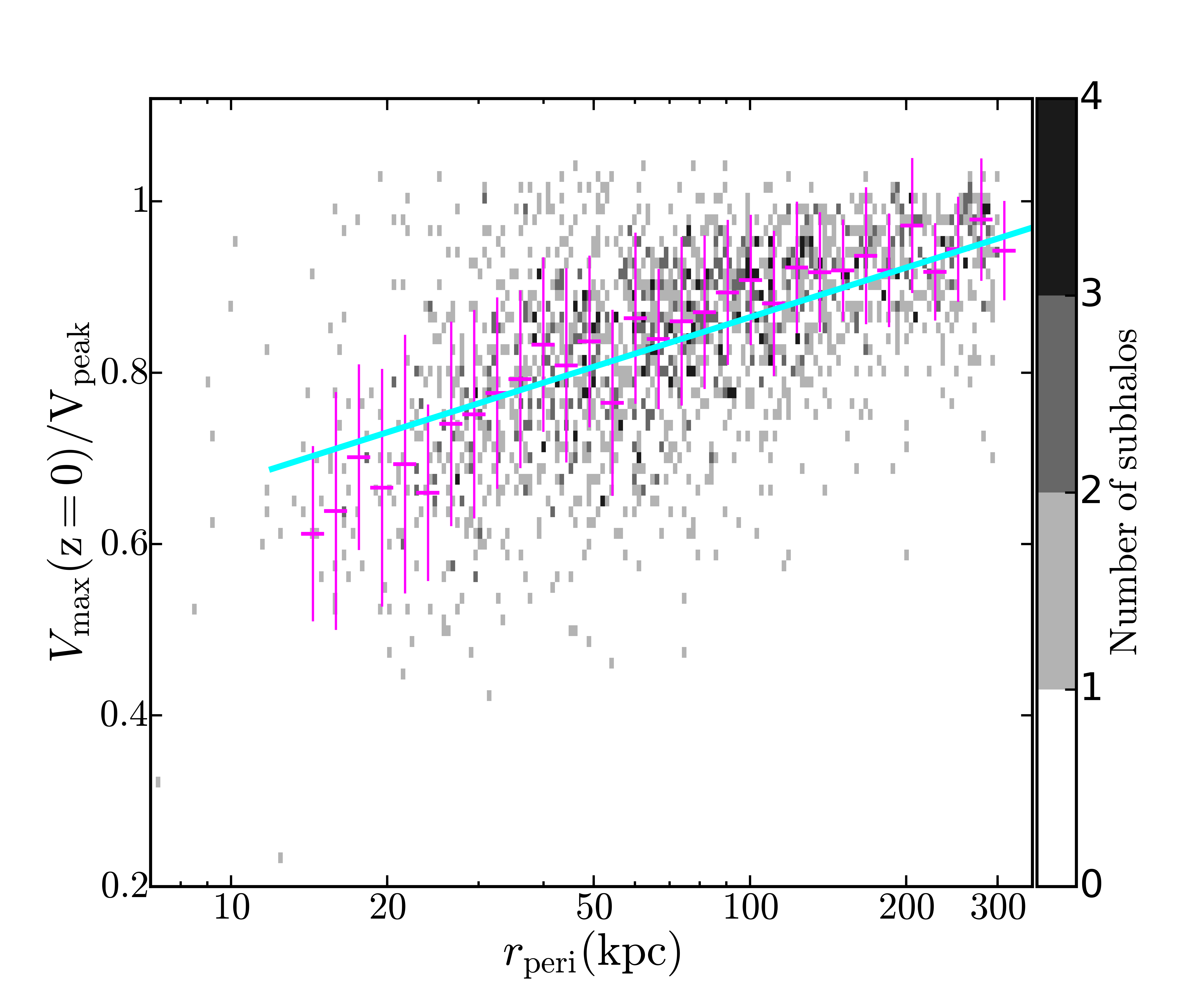} 

 \caption{
 2D-Histogram of the ratio between the maximum circular velocity at $z=0$ and its value prior to infall for all subhalos with $V_{\rm max} \geq 10 \rm km s^{-1}$ in the phat-ELVIS MW suite as a function of the pericenter distance from the center of their respective host. We show the median value of the simulation data with its 1-$\sigma$ scatter for each radial bin (magenta crosses) and our analytic best-fit relation describing the $V_{\rm max}(z=0)/V_{\rm peak}$-$r_{\rm peri}$ correlation (cyan line), which traces the median. This correlation indicates that subhalos undergo larger tidal stripping for decreasing pericenter distances. The smallest pericenters have current $V_{\rm max}$ values reduced by almost a factor of two compared to the value prior to infall.}
\end{figure}

\begin{figure*}
\centering
\includegraphics[width=17.cm, height=11.cm]{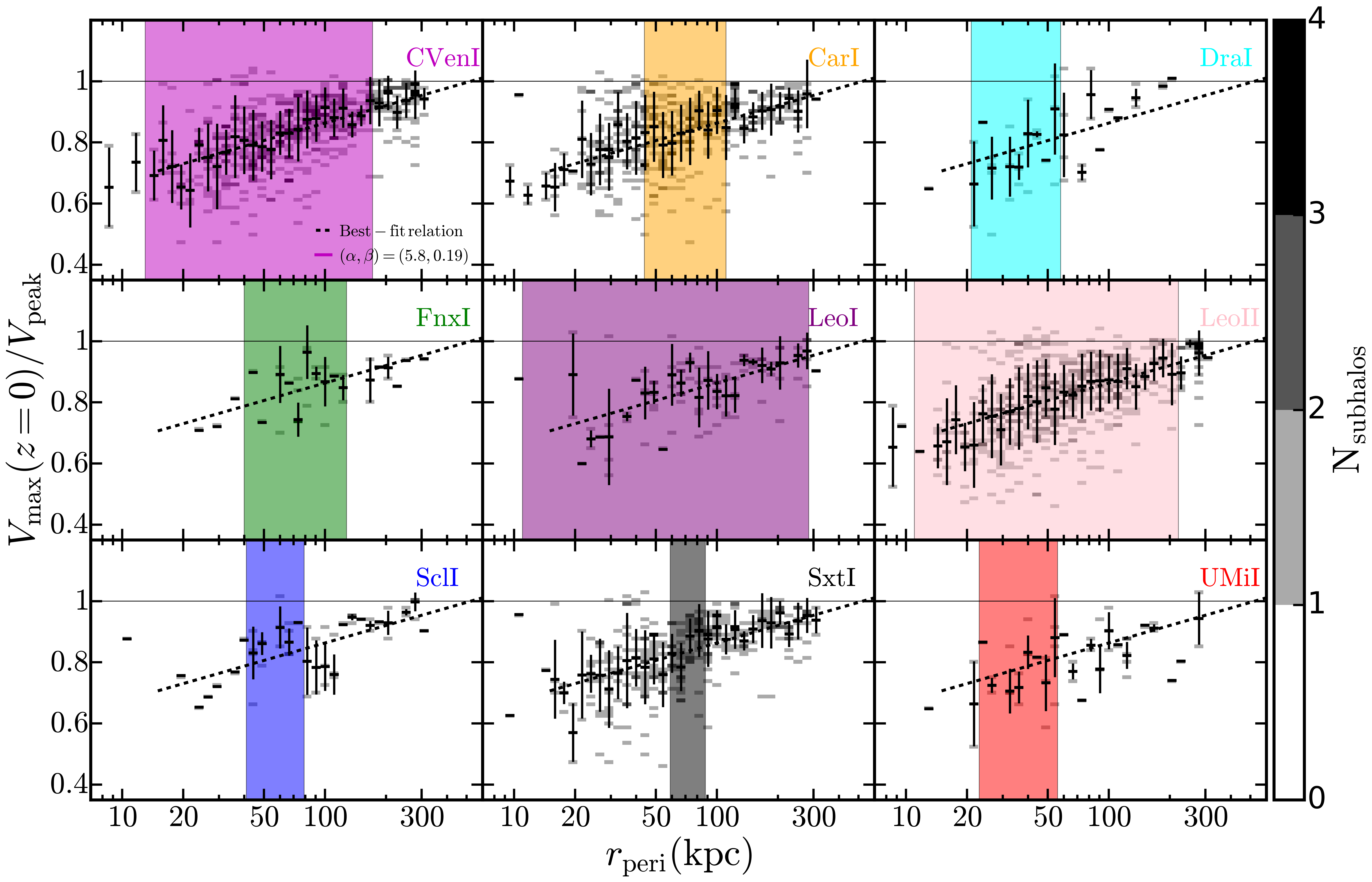}
\caption{  2D-histogram of the $V_{\rm max}-r_{\rm max}$ values for those subhalos that are consistent with the $M_{1/2},r_{1/2}$ values of the respective MW dSph labeled in each panel. The vertical regions are the constraints in pericenter distances obtained in \citet{Fritz18}. Subhalos that fall inside these regions are \textit{simultaneously} consistent with observed orbital pericenter constraints and internal dynamical constraints at $M_{1/2}$. The black points with error bars show the radially-binned median plus $1-\sigma$ scatter.  The dashed lines are the  best-fit relation for the entire sample, Eq. (\ref{eq:vrat})(dashed-line).  In each case the best-fit relation tracks well the respective median distribution.}
\end{figure*}

\section{Simulations and Data}\label{sec:sims}

We use data from the Phat-ELVIS suite \citep{kelley18} simulations, which model Milky Way-size  dark matter halos using dark-matter-only simulations coupled with  time-dependent embedded baryonic potentials that mimic the Milky Way (MW) galaxy. The suite consist of 12 high-resolution \textit{zoom-in} \citep{katz93,onorbe14} simulations of MW hosts, which span virial masses $M_{\rm vir} \approx 0.7-2 \times 10^{12} M_\odot$, each with an embedded galaxy potential grown to match the observed Milky Way disc and bulge today \citep[for details see][]{kelley18}). 
These simulations have a dark matter particle mass of $m_{\rm dm}= 3 \times 10^4 M_\odot$ and a Plummer equivalent force softening length of 37 pc.  \citet{kelley18} found that subhalo catalogs are complete down to down to a maximum circular velocity $V_{\rm max}>4.5 \rm kms^{-1}$ (or a total bound mass $M\simeq 5 \times 10^6 M_\odot$). 

For each simulation, we select all haloes within 300 kpc of the host's centre and with a present maximum circular velocity $V_{\rm max} \geq 10 \, \rm km s^{-1}$ ($M \gtrsim 10^8 M_\odot$).  With this selection we are studying systems with more than $\sim 3000$ particles each. 
We characterize a subhalo prior to infall into its host via $V_{\rm peak}$, which is defined as the $V_{\rm max}$ when the subhalo's mass is maximized across time. For each subhalo we its maximum circular velocity, $V_{\rm max}$, and the radius where the circular velocity reaches its maximum, $r_{\rm max}$ directly from the particle data. All other simulated subhalo quantities are taken from the halo catalogs and merger trees made public by \citet{kelley18}. 

To compare with MW dSphs, we note that these are dispersion-supported dark matter-dominated systems.  For such systems, the mass within the deprojected half-light radius $r_{1/2}$  (M$_{1/2}:=$M$_{\rm dSph}(r_{1/2})$) is well-constrained by line-of-sight velocity measurements \citep{wolf10}. Combining velocity dispersion measurements from \citet{wolf10} and the updated deprojected half-light radii from \citet{munoz18} assuming a Plummer profile for the stellar distributions we obtain M$_{1/2}$ for each dwarf~\footnote{our main conclusions hold for other choices for the stellar density profile \citep{munoz18}.}.

We derive constraints on each classical dSph $V_{\rm max}$ following \citep{boylan11}, who showed that the following approach provides the correct mass to better than 10 per cent at resolved radii.  We first assume that the current density distribution of a subhalo hosting a dwarf can be modeled by an \citet{nfw97} (hereinafter NFW) profile, which is a good approximation within $r_{\rm max}$ for typical CDM subhalos. For each of the nine classical dSphs we search for the combination of values $\{V_{\rm max},r_{\rm max}\}$ defining an NFW halo that would yield an enclosed subhalo mass consistent with the observationally-inferred dSph mass within 1-$\sigma$, i.e. that M$_{\rm sub}(r_{1/2})$ matches the observed M$_{1/2}$. 

\section{Results}\label{sec:results}

In Fig. \ref{fig1}, we show the range of $\{V_{\rm max},r_{\rm max}\}$ values that are consistent with the 1-$\sigma$ confidence region of $M_{1/2}$ for each of the  nine classical dSphs.
Each contour has a global minimum at $V_{\rm max}=\sqrt{3} \sigma_{\rm los,\star}$ corresponding to $r_{\rm max}$=$r_{1/2}$ \citep{wolf10}. 
Also shown are the $V_{\rm max}$-$r_{\rm max}$ values for subhalos in the phat-ELVIS suite that have resolved densities at $r_{\rm max}$ (we use the definition of convergence in CDM halos discussed in \citet{hopkins18} in which halo densities converge past the radius enclosing 200 particles). The points are color coded by each subhalo's pericenter radius $r_{\rm peri}$, and demonstrate a trend for smaller $r_{\rm peri}$ to have smaller $r_{\rm max}$ at fixed $V_{\rm max}$.
The trend is well-characterized by a power-law fit: $(r_{\rm max}/1\rm kpc)= A (V_{\rm max}/10 \, \rm kms^{-1})^{1.5}$.  
If we divide subhalos into those with $r_{\rm peri} < 50$ kpc and those with $r_{\rm peri} > 50$ kpc, we find that the normalization in the $V_{\rm max}$-$r_{\rm max}$ relation shifts from $A = 0.648$ to $A=0.893$ as we go from the small pericenter group to the larger pericenter group.  

The original TBTF problem \citep{boylan11} was framed using a plot similar to that in Fig. 1.  The  points that sit above the Draco (DraI)  and Ursa Minor (UMiI) bands are the problematic CDM subhalos.  These ``massive failures" (with $V_{\rm max} \gtrsim 30$ km s$^{-1}$) are too dense to host any known dwarf.  One proposed solution to the problem has involved enhanced depletion associated with a central galaxy potential \citep{zolotov12,Brooks_2014}.  
Here we see that even when depletion associated with a realistic MW baryonic potential is present, the issue is not necessarily solved, with many points sitting high in the plot.   As we will discuss below, the two densest dwarfs (UMiI and DraI) happen to also have observed pericenters that are fairly small ($\lesssim 50$ kpc). This exacerbates the problem: subhalos with smaller pericenters tend to over-populate problematic region above Draco and Ursa Minor in this plot.  That is, conditional on these galaxies having small pericenters, we would infer even lower $V_{\rm max}$ values ($\lesssim 25$ km s$^{-1}$) than may have been expected otherwise.  We return to this issue below.

When small halos are accretted into larger hosts and become subhalos, they begin to lose mass.  One measure of the degree of mass loss is the ratio  $V_{\rm max}/V_{\rm peak}$, where lower ratios imply more mass loss.  Typically, $V_{\rm peak}$ estimates for MW dSphs are inferred statistically based on their observationally-inferred $V_{\rm max}$ values and the expectation for the full $V_{\rm max}/V_{\rm peak}$ distribution from an entire subhalo population \citep[e.g.][]{boylan12}.   
In Fig. 2 we show that with if constraints on $r_{\rm peri}$ are available, this may allow us to narrow the inferred range.  Specifically, we show the
the ratio of $V_{\rm max}$ at present to its value at infall, $V_{\rm peak}$, for all selected subhalos in the 12 MW runs as a function of the subhalo's pericenter with respect to the host center. We find a robust trend of decreasing for smaller $r_{\rm peri}$ (Pearson correlation coefficient r=0.54). Though not shown, we find that this trend holds independent of the MW host virial mass in our sample and is independent of the precise choice of the minimum subhalo $V_{\rm max}$. We find the average trend is described by 
\begin{equation} \label{eq:vrat}
\frac{V_{\rm max}(z=0)}{V_{\rm peak}}= \log_{10} \bigg [ \alpha \bigg ( \frac{r_{\rm peri}}{30 \, \rm kpc} \bigg )^\beta \bigg ],
\end{equation}
with $\alpha=5.80$ and $\beta=0.19$ as best-fit parameters. Although more complex fitting functions are possible, this simple equation captures the systematic effect of increasing halo mass loss with pericenter distance. We find our fit is representative of the radially-binned median distribution (magenta points in Fig. 2) for $ 12 \kpc \lesssim r_{\rm peri} \leq 300 \kpc$.  This fit becomes a poor description below $\sim 12 \kpc$.  The detailed trend below this scale is difficult to discern because there are so few surviving subhalos with very small pericenter \citep{kelley18,robles19}.

For each of the nine dSph galaxies in the MW we show in Fig. 3 the relation between $V_{\rm max}/V_{\rm peak}$ and $r_{\rm peri}$ (obtained from the merger trees) of the subhalos (gray points) that could host the dSph (those overlapping with the respective shaded region in Fig. 2). We also show the best-fit relation Eq. (\ref{eq:vrat}), noting that our fit is representative of the radially-binned median distribution (black points) in all cases. 

The colored bands in Fig. 3 illustrate how observed orbital information may be used to tighten estimates of mass loss for each dwarf.  Specifically they reflect observational estimates of each dwarf galaxy's pericenter from \citet{Fritz18}.   These authors report $r_{\rm peri}$ values assuming two different MW virial masses ($0.8,1.6 \times 10^{12}M_\odot$).  These masses span the host mass ranges for the halos we use from the phat-ELVIS suite. We take the lowest and highest 1-$\sigma$ values for either of the virial masses assumed in \cite{Fritz18} to bracket the observational range. 
By imposing these new constraints, we identify a population of subhalos that are consistent with both internal mass estimates $M_{1/2}$ {\em and} inferred $r_{\rm peri}$ ranges for each dwarf.

\begin{figure}
\centering
\includegraphics[width=8.8cm, height=6.5cm]{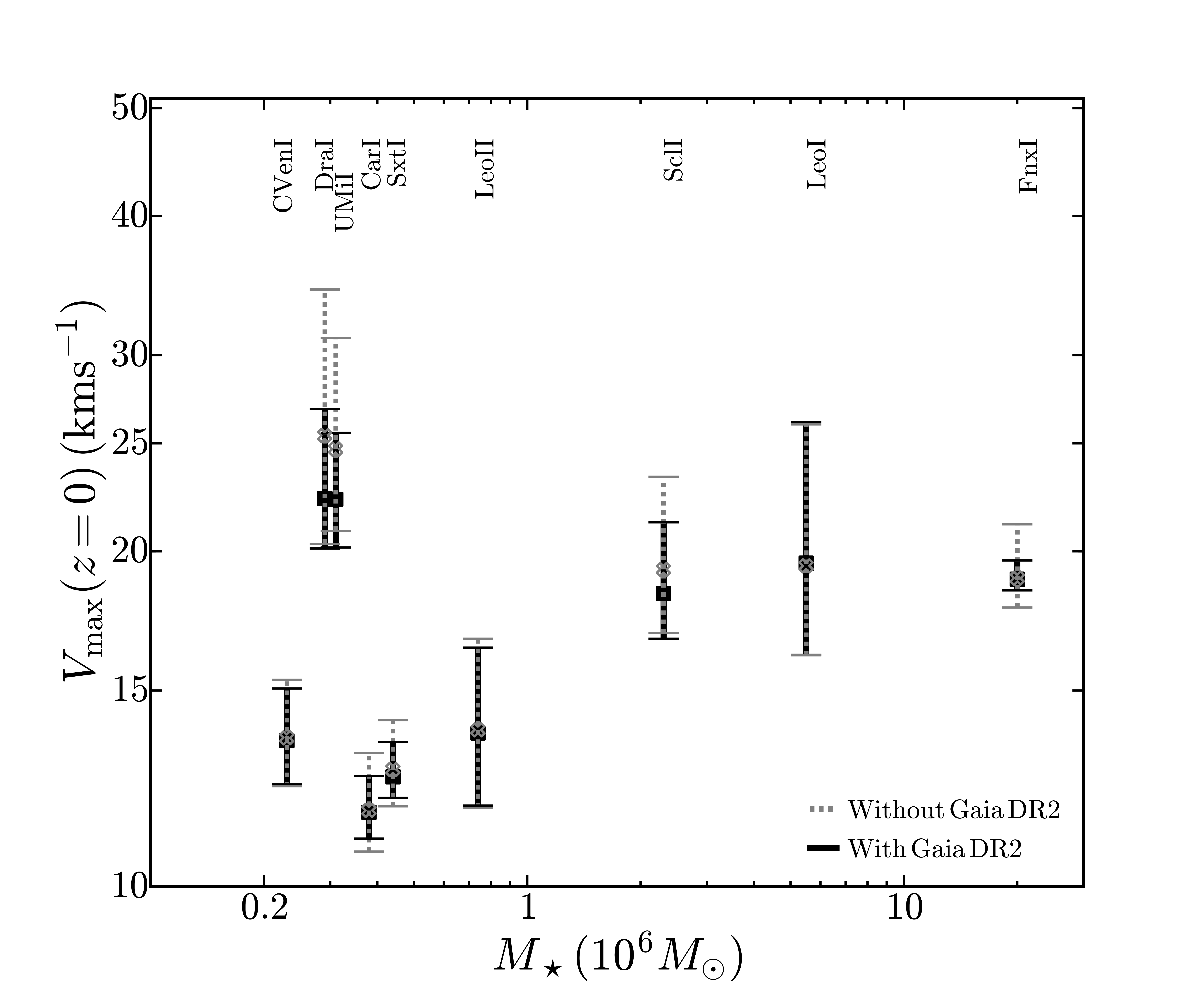} 
\includegraphics[width=7.5cm, height=6.5cm]{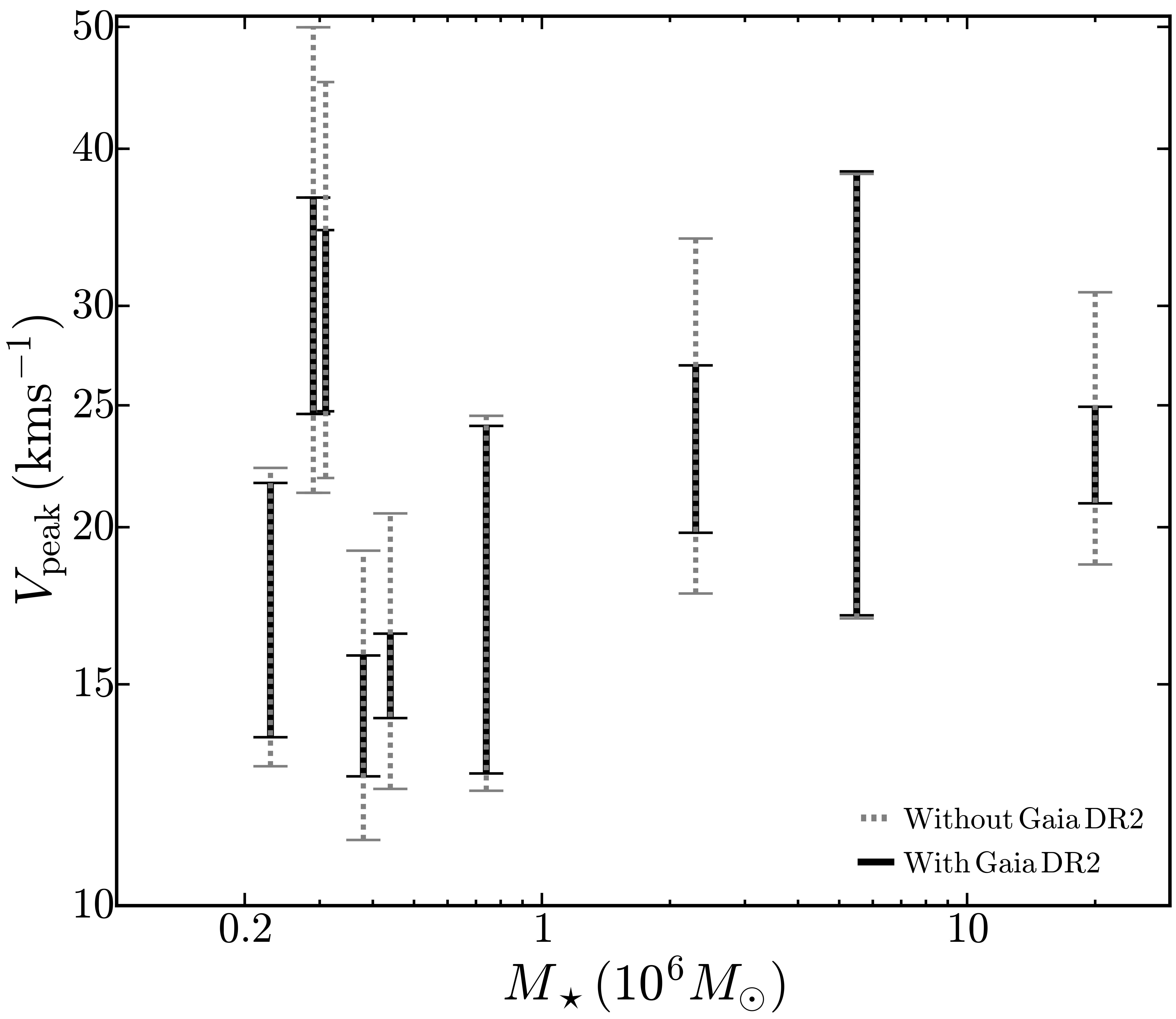}

 \caption{
 Constraints for $V_{\rm max}$ (top) and $V_{\rm peak}$ (bottom) for each the nine bright MW dwarf galaxies obtained from subhalos that satisfy the dynamical constraints (dashed lines) and those that in addition satisfy the pericenter constraints from \textit{Gaia}(solid lines). We show the measured stellar masses from \citet{mcon12}, we intentionally shifted the mass of Ursa Minor by $10^4M_\odot$ to visually distinguish it from Draco, we found that $V_{\rm max}$ and $V_{\rm peak}$ are not correlated with $M_\star$. For a given $V_{\rm max}$ there is a range of  $V_{\rm peak}$ due to the dependence of $r_{\rm peri}$ in Eq (\ref{eq:vrat}), in the bottom panel we show the full allowed range of $V_{\rm peak}$ inferred from our constraints of $V_{\rm max}$ within 1$\sigma$ (upper panel).}

\end{figure}

Table 1 presents $V_{\rm max}$  and $V_{\rm peak}$ ranges for each of the nine MW dSphs, inferred both with and without orbital pericenter constraints. We provide results for median, mean, and one sigma (1$\sigma$) ranges in each case.  Fig. 4 summarizes these results by plotting $V_{\rm max}$ (top panel) and $V_{\rm peak}$ (bottom panel) as a function of stellar mass \citep[taken from][]{mcon12}. In both panels, the solid black lines show $\pm$1$\sigma$ ranges when pericenter constraints are included and the dotted lines show how those ranges increase when pericenter constraints are neglected. In the top panel, the black and gray points are the inferred averages with and without pericenter constraints. In most cases, inferred $V_{\rm max}$ values shift lower by $20-80\%$, with somewhat smaller 1$\sigma$ ranges. The range of inferred $V_{\rm peak}$ values are similarly tighter. Exceptions are CVenI, LeoII, and LeoI, all of which have largely unconstrained pericenters ranges (see Fig. 3).

The most notable differences are for Draco and Ursa Minor, both of which have fairly well constrained pericenter measurements, $r_{\rm peri} \lesssim 50$ kpc. For Draco, the estimated range for $z=0$ maximum circular velocity shifts from $V_{\rm max} \simeq (20 - 43)$ to $(20 - 27) \: \rm km s^{-1}$.  For Ursa Minor,  $V_{\rm max} \simeq (20 - 31)$ narrows to $(20 - 26) \: \rm km s^{-1}$.  Their inferred $V_{\rm peak}$ ranges shift from $(21 - 50)$ to $(25 - 37) \: \rm km s^{-1}$ in the case of Draco and from $(21 - 50)$ to $(25 - 37) \: \rm km s^{-1}$ for Ursa Minor.

%%%%%%%%%%%%%%%%%%%%%% tab1 %%%%%%%%%%%%%%%%%%%%%
\begin{table*}
\setlength{\tabcolsep}{3pt}
\begin{tabular}{lccccccccc} % \toprule
\multicolumn{10}{c}{\ \ \ \ \ \  \ \ \ \ \ \ \ \ \ \ \ \ \ \ \ \ \ \ \ \ Without Gaia DR2 \ \ \ \ \ \ \ \ \ \ \ \ \ With Gaia DR2 \ \ \ \ \ \ \ \ \ \  \ Without Gaia DR2 \ \ \ \ \  With Gaia DR2} \\ 

Galaxy    & $r_{1/2}$ & $\widetilde{V}_{\rm max}(z=0)$             & $\bar{V}_{\rm max} \pm 1 \sigma$                     & $\widetilde{V}_{\rm max}(z=0)$  & $\bar{V}_{\rm max} \pm 1 \sigma$ & $V_{\rm peak,min}$& $V_{\rm peak,max}$ &$V_{\rm peak,min}$& $V_{\rm peak,max}$\\
               &$\left[\rm pc \right]$ & $\left[\rm km s^{-1} \right]$ & $\left[\rm km s^{-1} \right]$ & $\left[\rm km s^{-1} \right]$ & $\left[\rm km s^{-1} \right]$ &$\left[\rm km s^{-1} \right]$ & $\left[\rm km s^{-1} \right]$ & $\left[\rm km s^{-1} \right]$ & $\left[\rm km s^{-1} \right]$\\ 
          %\hline\hline

CVen I  & 589 & $13.59$ & 13.82 $\pm$ 1.51                         & 13.52                & 13.70 $\pm$ 1.35 & 12.90  & 22.29  & 13.61& 21.69\\
Carina I & 401 & $11.70$ & 11.96 $\pm$ 1.21                         & 11.65               &  11.80 $\pm$ 0.76 & 11.27 & 19.15 &12.67 & 15.81\\                   
Draco I & 279 & $25.40$ & 27.33 $\pm$ 7.02                         & 22.31                &  23.48 $\pm$ 3.36  & 21.30& 49.96 & 24.60& 36.58\\                  
Fornax I & 1093  & $18.91$ & 19.47 $\pm$ 1.67                        & 18.87               & 19.03 $\pm$ 0.58   & 18.67 & 30.75 & 20.89&24.93\\   
Leo I & 352 & $19.40$         & 21.06 $\pm$ 4.93                       & 19.51                & 21.13 $\pm$ 4.98   &16.92 & 38.21 & 17.02& 38.37\\   
Leo II & 223 & $13.82$         & 14.23 $\pm$ 2.46                       & 13.73               & 14.10 $\pm$ 2.28   & 12.34  & 24.53  & 12.73 & 24.07\\ 
Sculptor I  & 365  & $19.26$ & 20.11 $\pm$ 3.22                         & 18.33          & 18.96 $\pm$ 2.26   & 17.71 & 33.93& 19.79 & 26.90\\     
Sextans I  & 539  & $12.73$ & 12.95  $\pm$ 1.15                         & 12.54           & 12.74 $\pm$ 0.73  &12.38 & 20.50  & 14.09& 16.45\\  
Ursa Minor I &  531   & $24.70$ & 25.97  $\pm$ 5.10                         & 22.26       & 22.85 $\pm$ 2.69  & 21.88 & 45.19 & 24.73& 34.46\\  
\end{tabular}
\caption{Columns are: (1) galaxy name; (2) Galaxy 3D half-light radius from \citet{munoz18}; (3) median values of $V_{\rm max}$ at $z=0$ and (4) mean $V_{\rm max}$ at $z=0$ and 1-$\sigma$ uncertainties when we do not use constraints from Gaia pericenters; analogous quantities are given in cols. (5) and (6) when we further apply Gaia pericenter constraints to derive the median and mean $V_{\rm max}$ today. (7) $V_{\rm peak,min}$ and col. (8) $V_{\rm peak,max}$ are the lower and upper bounds of allowed values inferred from Eq. ({\ref{eq:vrat}}) using the 1-$\sigma$ lower and upper bounds for $\bar{V}_{\rm max}$ when Gaia pericenter constraints are not considered,  columns (9) and (10) show the bounds when Gaia pericenter information is included.}
\label{tab:sims}
\end{table*}
%%%%%%%%%%%%%%%%%%%%%%%%%%%%%%%%%%%%%%%%%%%%%%%%%

\section{Summary and Discussion}

We have used phat-ELVIS suite of Milky Way-size halo simulations to show that subhalo orbital pericenters, $r_{\rm peri}$, correlate with internal dark matter halo structural properties and that this provides an avenue for inferring tighter constraints on satellite galaxies $V_{\rm vmax}$ and $V_{\rm peak}$ values. As illustrated in Figures 1 and 2, subhalos with smaller pericenters tend to be more concentrated, with smaller $r_{\rm max}$ values at fixed $V_{\rm max}$, and also have larger $V_{\rm peak}$ values at fixed $V_{\rm max}$. We then used published pericenter radii for the nine classical MW dSph galaxies from \citep{Fritz18} to infer tighter estimates on their maximum circular velocities today and prior infall (see Fig. 4 and Table 1). Even with these tighter constraints on satellite circular velocites, we see no indication of a trend between stellar mass and halo mass at these scales.

Among the galaxies we have considered, Draco and Ursa Minor are affected the most by including pericenter information.  These, the densest dSph satellites, have relatively small pericenters ($r_{\rm peri} \lesssim 50$ kpc) and this pushes their inferred  $V_{\rm max}$ values lower than they would have been without pericenter information.   We find that both Draco and Ursa Minor appear to inhabit subhalos with $V_{\rm max} \simeq 20 - 27$ km s$^{-1}$ (at one sigma) and are less dense than would be expected for the most massive subhalos with small pericenters (see Fig. 1).  This exacerbates the traditional TBTF problem \citep{boylan11,boylan12}. The inferred range of $V_{\rm peak}$ values is also tightened for these galaxies.  Both galaxies have inferred peak maximum circular velocities (prior to infall) that put them in range to have been significantly affected by reionization, $V_{\rm peak} \simeq 25 - 37$ km s$^{-1}$ \citep[e.g.][]{gnedin2000,bullock2000,fitts17,bl2017}.

It is worth considering whether our results are affected by numerical resolution.  There is a fair amount of evidence to suggest that simulations such as ours do not accurately track halo survival after they have lost a significant ($\gtrsim 99\%$) amount of mass, with $\sim 3000$ particles appearing to be a critical limit below which halos disrupt unphysically \citep[e.g.][]{vandenbosch18,Errani2020}.   In terms of halo counts alone, this effect should not affect current cosmological zoom simulations significantly, since accreted halo mass functions rise steeply with increasing mass, $dn/dM \sim M^{-1.8}$ \citep[e.g.][]{Giocoli08}. This means that for every halo of mass $M$ that was accreted long ago and has lost 99\% of its original mass (100 $M$), there are as many as $\sim 4000$ halos of mass $M$ that were accreted more recently and that can be accurately counted. As discussed in \citet{kelley18}, convergence tests suggest that our halo counts are mostly complete down to $V_{\rm max} > 4.5$ km s$^{-1}$, or only $\sim 150$ particles. Of particular relevance to this work is whether or not internal halo structures, e.g., $V_{\rm max}/V_{\rm peak}$ distributions at fixed $V_{\rm max}$, are tracked appropriately. We have limited our analysis to halos with $V_{\rm max} > 10$ km s$^{-1}$, or $\gtrsim 3000$ particles. According to the results of \citet{Errani2020}, these should be fairly well resolved.  For Draco and Ursa Minor, we are looking at halos with $V_{\rm max} > 20$ km s$^{-1}$, which have $\gtrsim 40,00$ particles, so quoted results for these interesting systems are likely robust.  If anything, we will be biased towards missing halos with lower $V_{\rm max}/V_{\rm peak}$ values at small pericenter.  For example,  $V_{\rm max}/V_{\rm peak} \lesssim  0.3$ would imply $\gtrsim 99 \%$ mass loss. In the unlikely event that this this bias is significant, it would mean that the measured trend would become even stronger at higher resolution.  This will be a topic to explore in future simulations.

Our results appear consistent with the anti-correlation observed between the dark matter densities and pericenter radii of bright MW dwarfs reported by \citet{kaplinghat19}. In their work, the authors found that dwarf galaxies closer to the center of the MW tend to be hosted by denser CDM subhalos. We find that at fixed $V_{\rm max}$ subhalos with small pericenters are indeed denser (smaller $r_{\rm max}$, see Fig.1), as seems to be the case for MW satellites.   At fixed $V_{\rm peak}$ we find a similar trend (not shown).  An alternative explanation that might drive an even stronger anti-correlation may be that dark matter is self-interacting, which could drive core-collapse in high concentration low-mass subhalos \citep{hiroya20}. This was not observed in the analogous phat-ELVIS SIDM MW simulation in \cite{robles19} for a velocity-independent cross-section over particle mass $\sigma/m=1\rm cm^2/\rm g$.   More simulations are required to test the statistically significance of the abundance of core-collapsing halos and whether a larger $\sigma/m$ is needed.

As proper motion measurements begin to improve, there is hope that we will get a better sense of both the 3D orbital motions of MW satellite galaxies, but also their internal 3D velocity dispersion \citep{Massari_2018,Massari_2020}. These internal motions should provide tighter constraints on dark matter halo structure directly \citep{lazar20}. Our results suggest that improved orbital trajectories will only increase the power of these internal mass measurements to infer dark matter halo properties, both past and present.

\section*{Acknowledgments}

VHR acknowledges support by the Yale Center for Astronomy and Astrophysics postdoctoral prize fellowship and the Gary A. McCue postdoctoral fellowship. JSB and VHR were supported by NSF AST-1518291, HST-AR-14282, and HST-AR-13888.

\bibliography{mybib}

%%%%%%%%%%%%%%%%%%%%%%%%%%%%%%%%%%%%%%%%%%%%%%%%%%%%%%%%%%%%%%%%%
% \vfill\eject
\bsp
\label{lastpage}
\end{document}